\documentclass[reprint,superscriptaddress,bibnotes,amsmath,amssymb,aps,prl,floatfix]{revtex4-1}

\usepackage{graphicx}
\usepackage{dcolumn}
\usepackage{bm}
\usepackage{upgreek}
\usepackage{physics}
\usepackage{hyperref}

\begin{document}

\title{Light Scattering from Solid-State Quantum Emitters: Beyond the Atomic Picture}

\author{Alistair J. Brash}
\email[Email: ]{a.brash@sheffield.ac.uk}
\affiliation{Department of Physics and Astronomy, University of Sheffield, Sheffield, S3 7RH, United Kingdom}
\author{Jake Iles-Smith}
\email[Email: ]{Jakeilessmith@gmail.com}
\affiliation{Department of Physics and Astronomy, University of Sheffield, Sheffield, S3 7RH, United Kingdom}
\affiliation{School of Physics and Astronomy, The University of Manchester, Oxford Road, Manchester M13 9PL, UK}
\author{Catherine L. Phillips}
\affiliation{Department of Physics and Astronomy, University of Sheffield, Sheffield, S3 7RH, United Kingdom}
\author{Dara P. S. McCutcheon}
\affiliation{Quantum Engineering Technology Labs, H. H. Wills Physics Laboratory and Department of Electrical and Electronic Engineering, University of Bristol, Bristol BS8 1FD, UK}
\author{John O'Hara}
\affiliation{Department of Physics and Astronomy, University of Sheffield, Sheffield, S3 7RH, United Kingdom}
\author{Edmund Clarke}
\affiliation{EPSRC National Epitaxy Facility, Department of Electronic and Electrical Engineering, University of Sheffield, Sheffield, UK}
\author{Benjamin Royall}
\affiliation{Department of Physics and Astronomy, University of Sheffield, Sheffield, S3 7RH, United Kingdom}
\author{Jesper M\o{}rk}
\affiliation{Department of Photonics Engineering, DTU Fotonik, Technical University of Denmark, Building 343, 2800 Kongens Lyngby, Denmark}
\author{Maurice S. Skolnick}
\affiliation{Department of Physics and Astronomy, University of Sheffield, Sheffield, S3 7RH, United Kingdom}
\author{A. Mark Fox}
\affiliation{Department of Physics and Astronomy, University of Sheffield, Sheffield, S3 7RH, United Kingdom}
\author{Ahsan Nazir}
\affiliation{School of Physics and Astronomy, The University of Manchester, Oxford Road, Manchester M13 9PL, UK}
\date{\today}

\begin{abstract}
Coherent scattering of light by a single quantum emitter is a fundamental process at the heart of many proposed quantum technologies. Unlike atomic systems, solid-state emitters couple to their host lattice by phonons. Using a quantum dot in an optical nanocavity, we resolve these interactions in both time and frequency domains, going beyond the atomic picture to develop a comprehensive model of light scattering from solid-state emitters. We find that even in the presence of a cavity, phonon coupling leads to a sideband that is completely insensitive to excitation conditions, and to a non-monotonic relationship between laser detuning and coherent fraction, both major deviations from atom-like behaviour.
\end{abstract}

\maketitle


Scattering of light by a single quantum emitter is one of the fundamental processes of quantum optics. First observed in atomic systems~\cite{GIBBS197687,Volz2007}, and more recently studied extensively in self-assembled quantum dots (QDs)~\cite{Nguyen2011,PhysRevLett.108.093602,Proux2015,Bennett2016}, coherent scattering attracts interest as the scattered light retains the coherence of the laser rather than the emitter.
As such, their coherence may exceed the conventional radiative limit whilst still exhibiting antibunching on the timescale of the emitter lifetime~\cite{Nguyen2011,PhysRevLett.108.093602,Proux2015,Bennett2016}. Exploiting this behaviour gives rise to exciting possibilities for quantum technologies such as generating tuneable single photons~\cite{Matthiesen2013,PhysRevLett.111.237403,Sweeney2014}, realising single photon non-linearities~\cite{Javadi2015,Sipahigilaah6875,Bennett2016b,DeSantis2017,Hallett:18}, and constructing entangled states between photonic~\cite{PhysRevA.96.062329,PhysRevA.98.022318} or spin~\cite{Delteil2015,PhysRevLett.119.010503} degrees of freedom.

For a continuously driven emitter, coherent scattering occurs in the weak excitation regime where absorption and emission become a single coherent event.
For a simple two-level \emph{``atomic picture"} with only spontaneous emission and pure dephasing,
the coherently scattered fraction ($\mathcal{F}_{CS}$) of the total emission is given by \cite{*[{Eq. \ref{eq:CoherentFracRabi} generalises expressions from }][{ to include pure dephasing.}] CohenTannoudhiAtom}: 
\begin{equation}
    \mathcal{F}_{CS} 
    =\frac{T_2}{2T_1}\frac{1}{1+\mathcal{S}},
    \label{eq:CoherentFracRabi}
\end{equation}
where $\mathcal{S} = (\Omega^2 T_1 T_2) / (1 + \Delta_{LX}^2 T_2^2)$ is a generalized saturation parameter, $\Omega$ is the Rabi frequency, $\Delta_{LX}= \omega_{L} - \omega_{X}$ is the detuning between the laser ($\omega_L$) and the emitter ($\omega_X$) and $T_1$ and $T_2$ are the emitter life- and coherence times respectively.
It is clear from this expression that the fraction of coherently scattered light reaches unity in the limit of driving well below saturation ($\mathcal{S} \ll 1$) and transform-limited coherence ($T_2 = 2T_1$).

Solid-state emitters (SSEs), particularly self-assembled QDs, are an attractive system with which to realise such schemes owing to their high brightness and ease of integration with nanophotonic structures. 
However, unlike atoms, SSEs can experience significant dephasing from fluctuating charges~\cite{PhysRevLett.108.107401,Kuhlmann2013} and coupling to vibrational modes of the host  material~\cite{PhysRevLett.93.237401,PhysRevLett.118.233602}. 
Despite this, state-of-the-art InGaAs QD single photon sources have demonstrated essentially transform-limited photons emitted via the zero phonon line (ZPL)~\cite{Somaschi2016,PhysRevLett.116.213601,Liu2018} through careful sample optimisation, exploitation of photonic structures and by using resonant $\pi$-pulse excitation at cryogenic temperatures. 
Although these results show ZPL broadening can be effectively suppressed, coupling to vibrational modes also leads to the emergence of a phonon sideband (PSB) in the emission spectrum~\cite{PhysRevB.65.195313,Kaer2014,mccutcheon16,PhysRevB.95.201305,Iles-Smith2017,PhysRevLett.118.233602}.
This is attributed to a rapid change in lattice configuration of the host material during exciton recombination, leading to the simultaneous emission or absorption of longitudinal acoustic (LA) phonons with the emission of a photon~\cite{Besombes01,PhysRevB.65.195313,Kaer2014}. 
Therefore, to obtain perfectly indistinguishable photons the PSB must be filtered out, naturally limiting the efficiency of the device, even when using an optical cavity to Purcell enhance emission into the ZPL~\cite{Iles-Smith2017,PhysRevLett.118.253602}. 

\begin{figure*}
	\centering
	\includegraphics{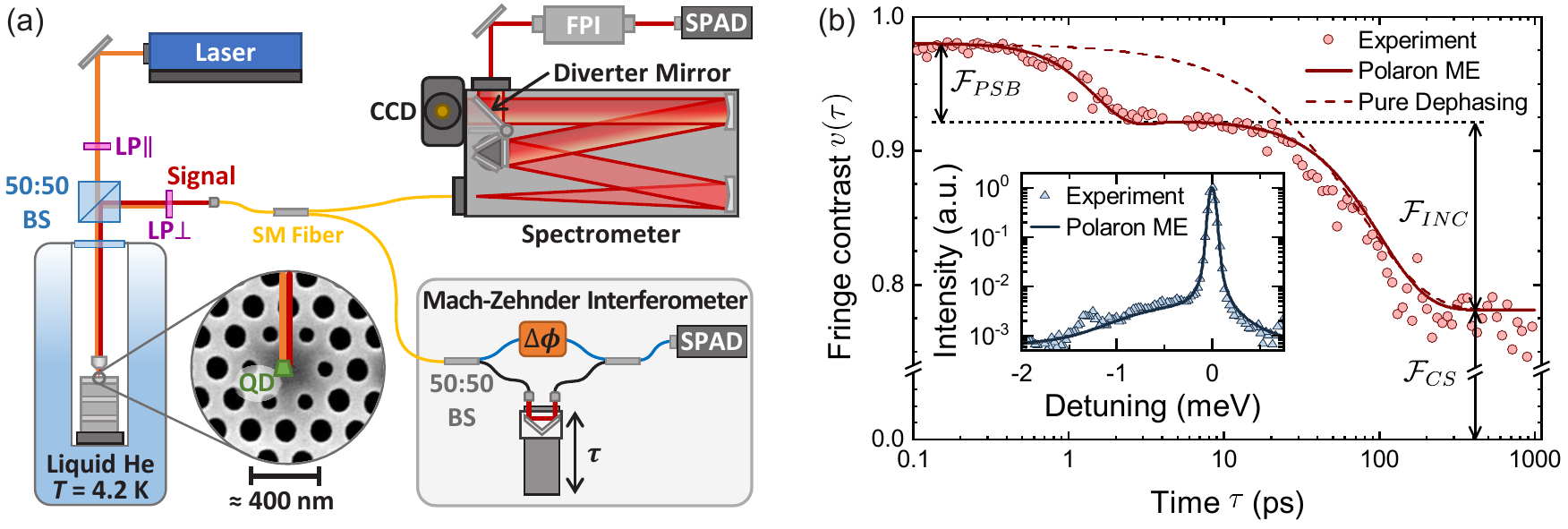}
	\caption{
	(a) Schematic of the experiment: BS - beam splitter, CCD - charge-coupled device (camera), FPI - Fabry-Perot interferometer, LP - linear polarizer, 
	SM - single mode fiber, SPAD - single photon avalanche diode, $\Delta \phi$ - phase shift, $\tau$ - path length difference.
	(b) Measurement of the first-order correlation function ($g^{(1)}(\tau)$) at $\mathcal{S}=0.25$ with $\Delta_{LX}=0$. The emission contains phonon sideband ($\mathcal{F}_{PSB}$), incoherent resonance fluorescence ($\mathcal{F}_{INC}$) and coherently scattered ($\mathcal{F}_{CS}$) fractions. Experimental measurements of fringe contrast (red circles) agree well with a calculation using the polaron master equation (solid red line) where the phonon coupling strength $\alpha$ and cut-off frequency $\nu_c$ are the only free parameters. A pure dephasing model (dashed red line) decays mono-exponentially and cannot capture phonon dynamics. \emph{Inset:} An experimental spectrum (blue triangles) measured simultaneously is also well reproduced by the polaron model (blue line) with the same parameters. The calculated spectrum is convolved with the spectrometer instrument response in order to reproduce the observed ZPL width.
	}
	\label{fig:PhononG1}
\end{figure*}

The aforementioned works~\cite{PhysRevB.65.195313,Kaer2014,mccutcheon16,PhysRevB.95.201305,Iles-Smith2017,PhysRevLett.118.233602,PhysRevLett.118.253602} have revealed the importance of phonon coupling in the incoherent regime, where there is a definite change of charge configuration in the QD, such that incoherently scattered resonance fluorescence dominates the spectrum.
It is perhaps natural to presume that phonon coupling may be eliminated by operating in the coherent scattering regime, since there is vanishing exciton population and therefore no change in charge configuration. 
This suggests that, in accordance with most works in the literature \cite{Nguyen2011,PhysRevLett.108.093602,Proux2015}, one may adopt the atom-like picture of Eq. \ref{eq:CoherentFracRabi}, where the coherent fraction tends towards unity for excitation far below saturation and transform-limited coherence.  
However, a recent theoretical study predicted that PSBs occur even for vanishingly weak resonant driving \cite{PhysRevB.95.201305}. 

Here, we experimentally verify that PSBs persist in the coherent scattering regime and demonstrate additionally that phonon processes also cause large deviations from atom-like physics when driving off-resonance.
An extended theoretical model fully describes our solid-state nanocavity system, 
providing an intuitive picture that attributes the PSB to phonon dressing of the optical dipole moment.
This leads to a finite probability that the vibrational environment changes state during an optical scattering event, implying that all optical spectral features will have an associated PSB. 
Whilst a self-assembled QD is studied here, we emphasize that the physics and methods apply equally to a diverse range of SSEs, including vacancy centers in diamond~\cite{Faraon2011,1367-2630-13-2-025012}, defects in hexagonal boron nitride \cite{PhysRevLett.119.057401}, monolayer transition metal dichalcogenides \cite{PhysRevLett.119.187402} and single carbon nanotubes \cite{PhysRevLett.116.247402}. 

To study this phonon coupling experimentally, we investigate a neutral exciton state ($\ket{X}$) of a self-assembled InGaAs QD with dipole moment $|\vec{\mu}|=27.2~\mathrm{D}$, weakly coupled ($\hbar g = 135 ~\upmu \mathrm{eV}$) to a H1 photonic crystal cavity (linewidth $\hbar \kappa = 2.51~\mathrm{meV}$) with Purcell factor $F_{P}=43$ (see Ref.~\cite{Liu2018} for full details). 
As well as Purcell enhancing the ZPL, the cavity also acts as a weak spectral filter; this combination can reduce the PSB component of the emission~\cite{Iles-Smith2017,PhysRevLett.118.253602}, motivating the coupling of SSEs to cavities.  
Fig. \ref{fig:PhononG1}(a) illustrates the experiment; the sample is held in a liquid helium bath cryostat at $T = 4.2~\mathrm{K}$ and excited by a tuneable laser that is rejected from the detection path by cross-polarisation (typical signal-to-background $>$100:1). The coherence of the scattered light is studied either in the time domain by measuring the fringe contrast $v(\tau)$ in a Mach-Zehnder interferometer or in the frequency domain using a spectrometer or a Fabry-Perot interferometer (FPI) for higher resolution (details in supplemental material (SM)~\cite{Sup}).

It is instructive to begin with a high resolution time-domain measurement, exciting resonantly below saturation ($\mathcal{S} = 0.25$) where coherent scattering is expected to dominate the emission. The measured fringe contrast $v(\tau)$ is proportional to the first order correlation function $g^{(1)}(\tau)$~\cite{Sup}. 
The result in Fig. \ref{fig:PhononG1}(b) departs significantly from the mono-exponential radiative decay predicted by atomic theory (dashed line);
a rapid decay of coherence occurs in the first few picoseconds, comparable to phonon dynamics observed in pulsed four-wave mixing measurements of InGaAs QDs in the incoherent regime \cite{doi:10.1021/acsphotonics.6b00707}, suggesting that the rapid loss of coherence we observe originates from electron-phonon interactions. 
 
In order to describe such behaviour accurately, we must account for the microscopic nature of the QD-phonon coupling~\cite{Nazir_Review_2016}.
This is achieved by applying the polaron transformation to the full system-environment Hamiltonian, dressing the excitonic states of the system with modes of the phonon environment. 
We may then derive a master equation (ME) that is non-perturbative in the electron-phonon coupling strength~\cite{breuer2002theory,McCutcheon_2010,PhysRevLett.106.247403,PhysRevB.95.201305} to describe the evolution of the reduced state of the QD~\cite{Sup}.
In the polaron frame, the first-order correlation function is 
$g^{(1)}_\mathrm{pol}(\tau) = \mathcal{G}(\tau)g^{(1)}_{opt}(\tau)$~\cite{PhysRevB.95.201305}, where $g^{(1)}_{opt}(\tau)$ is the purely optical contribution found using the polaron frame ME, while $\mathcal{G}(\tau) = B^2 \exp(\varphi(\tau))$ is the correlation function of the phonon environment, which accounts for non-Markovian phonon relaxation.
Here we have defined the phonon propagator $\varphi(\tau) = \alpha\int_0^\infty \nu e^{-\nu^2/\nu_c^2}(\cos(\nu\tau)\coth(\nu/2k_BT) - i\sin(\nu\tau))d\nu$, and the Franck-Condon factor $B = \exp(-\varphi(0)/2)$.
The coupling of the QD to the phonon environment is thus specified by its thermal energy $k_BT$, the deformation potential coupling strength $\alpha$, and cut-off frequency $\nu_c$~\cite{PhysRevB.65.195313,PhysRevB.84.125304,Nazir_Review_2016}. 
The cavity leads both to Purcell enhancement of the exciton transition (included within the ME) and spectral filtering of the emission. 
We incorporate cavity filtering by solving the Heisenberg equations of motion for the cavity field operators, leading to the detected function: 
\begin{equation}
g^{(1)}_{D}(\tau) = \int_{-\infty}^\infty \tilde{h}(t-\tau)g^{(1)}_{\mathrm{pol}} (t)\,dt,
\label{eq:G1D}
\end{equation}
where $\tilde{h}(t) = \exp(-i\Delta_{XC}t - \kappa\vert t\vert/2)$ is the cavity filter function and $\Delta_{XC}$ is the exciton-cavity detuning~\cite{Sup}.

By fitting the phonon bath correlation function contained within Eq. \ref{eq:G1D} to the first few picoseconds of the measurement, we extract phonon parameters 
$\alpha = 0.0447~\mathrm{ps^2}$ and 
$\nu_c = 1.28~\mathrm{ps^{-1}}$, comparable to values previously found for InGaAs QDs \cite{PhysRevLett.104.017402}. 
Using experimentally determined values for all other parameters,  
we accurately reproduce the full dynamics of the experimental data, as shown by the solid line in Fig. \ref{fig:PhononG1}.
After phonon relaxation, radiative decay associated with incoherent resonance fluorescence occurs between $\tau=20-200~\mathrm{ps}$. 
Finally, at $\tau \gg 200~\mathrm{ps}$, $v(\tau)$ plateaus, corresponding to the coherent fraction of the emission. As the laser coherence time is much greater than the measured delays, no decay of the coherent scattering is observed. 
From the $v(\tau)$ amplitudes, we extract $\mathcal{F}_\mathrm{PSB} = 0.06$, $\mathcal{F}_\mathrm{INC} = 0.14$ and $\mathcal{F}_\mathrm{CS} = 0.80$ respectively for the PSB, incoherent and coherent fractions of the total emission ($\mathcal{F}$). Crucially, a finite $\mathcal{F}_\mathrm{PSB}$ under weak driving indicates that Eq. \ref{eq:CoherentFracRabi} does not fully describe the scattering dynamics of the system.

To check the accuracy of the extracted parameters we now move to the frequency domain.
The theoretical spectrum is calculated by Fourier transforming $g^{(1)}_D(\tau)$ and may be written as: 
$
   S(\omega) = H(\omega)(S_\mathrm{opt}(\omega) + S_\mathrm{SB}(\omega)), 
$ 
where $H(\omega) = (\kappa/2)/[(\omega-\Delta_{XC})^2 + (\kappa/2)^2]$ is the frequency domain cavity filter function~\cite{RoyChoudry15,Iles-Smith2017,denning2018cavity}.
The spectrum consists of two principal components: a purely optical part
\begin{equation}
S_\mathrm{opt}(\omega) = B^2\int_{-\infty}^\infty g_\mathrm{opt}(\tau)e^{i\omega\tau}d\tau,
\label{eq:G1Opt}
\end{equation}
containing both coherent and incoherent contributions to the spectrum, and a second incoherent component
\begin{equation}
 S_\mathrm{SB}(\omega)= \int_{-\infty}^\infty \left(\mathcal{G}(\tau)-B^2\right)g_\mathrm{opt}(\tau)e^{i\omega\tau}d\tau,
 \label{eq:G1SB}
\end{equation}
which gives rise to the PSB~\cite{PhysRevB.95.201305, Iles-Smith2017}. 
The ZPL contribution is thus reduced by the square of the constant Franck-Condon factor $B^2$, with the missing fraction emitted through the PSB.

The inset to Fig. \ref{fig:PhononG1} illustrates that the parameters extracted from the time domain dynamics lead to excellent agreement between the experimental (blue triangles) and theoretical ($S(\omega)$ - solid line) spectra, with a broad PSB observed in accordance with the short timescale of the phonon processes. 
These combined time and frequency domain measurements provide critical insight into the nature of electron-phonon interactions in driven QDs: even well below saturation, where the excited state population is small and coherent scattering dominates, a PSB is present, comprising $\sim6$\% of the emission.

\begin{figure*}
	\centering
	\includegraphics{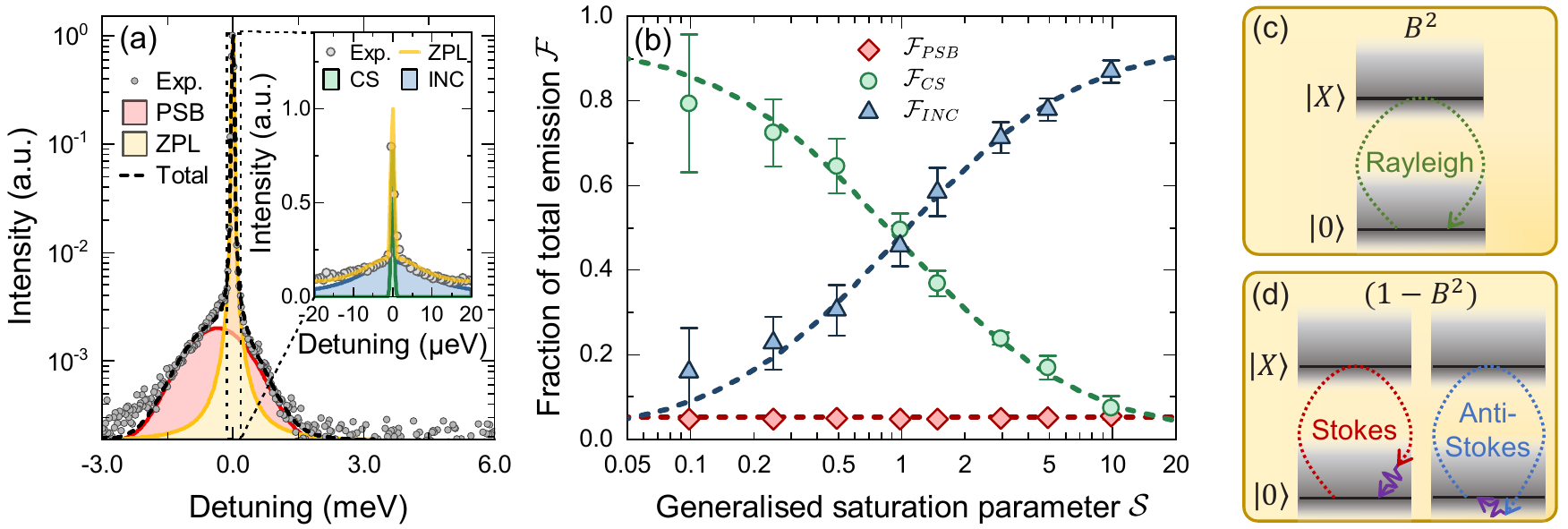}
	\caption{Components of the resonant scattering spectrum. (a) Semi-log spectrum at low resolution ($\mathcal{S}=10$, $\Delta_{LX}=0$): grey circles - experiment; yellow - fit to ZPL; red -  fit to PSB; black dashed line - total fit.  \emph{Inset:} High resolution spectrum: grey circles - experiment; green - fit to coherent scattering; blue - fit to incoherent resonance fluorescence; yellow - total fitted profile. (b) Evolution with increasing $\Omega$: red diamonds - PSB; green circles - coherent scattering; blue triangles - incoherent resonance fluorescence; dashed lines - polaron model. (c) Coherent scattering occurs when laser photons scatter directly from the bare transition (solid black lines) with probability given by the square of the Franck-Condon factor ($B^2$). (d) Inelastic scattering occurs when the system scatters (with probability $(1-B^2)$) into a different vibrational state within the ground-state manifold (grey shading). An LA phonon (purple) is emitted or absorbed for the Stokes and anti-Stokes cases respectively.}
	\label{fig:PSBSpectraRabi}
\end{figure*}

To investigate to what extent the PSB persists in the coherent scattering regime, we measure the resonance fluorescence spectrum as a function of saturation by varying $\Omega$. 
Fig. \ref{fig:PSBSpectraRabi}(a) shows a spectrum taken well above saturation ($\mathcal{S}=10$) that exhibits a ZPL (yellow fit) and a PSB ($S_{SB}(\omega)$ - red fit). 
Performing high resolution spectroscopy of the ZPL with the FPI results in the inset to Fig. \ref{fig:PSBSpectraRabi}(a) which exhibits a broad contribution from incoherent resonance fluorescence (blue fit) and a narrow feature from coherent scattering (green fit). 
As in the ${g}^{(1)}(\tau)$ of Fig. \ref{fig:PhononG1}(b), the total spectrum thus comprises three components whose fraction of the total emission can be evaluated from their areas (details in \cite{Sup}).

Fig. \ref{fig:PSBSpectraRabi}(b) shows the evolution of the components of the resonant
($\Delta_{LX}=0$) scattering spectrum as a function of $\mathcal{S}$. 
The polaron model agrees well with the experiment and produces a curve for $\mathcal{F}_{CS}$ (green dashed line) that is proportional to $(1+\mathcal{S})^{-1}$ like Eq. \ref{eq:CoherentFracRabi}. 
However, as previously predicted~\cite{PhysRevB.95.201305}, $\mathcal{F}_{CS}$ does not reach unity for vanishing $\mathcal{S}$, a surprising result that may be explained by observing that the PSB fraction $\mathcal{F}_{PSB}$ (red diamonds) is constant and independent of $\Omega$. This contrasts with excitation induced dephasing (EID)~\cite{PhysRevLett.104.017402,Ulhaq:13} which is also mediated by LA phonons and captured within our model, but is proportional to $(\Omega^2+\Delta_{LX}^2)$ and is thus negligible for resonant driving below saturation.

The results of Fig. \ref{fig:PSBSpectraRabi}(b) can be understood by considering
the possible scattering channels illustrated in Figs. \ref{fig:PSBSpectraRabi}(c,d). 
The bare transition $\ket{0} \rightarrow \ket{X}$ (solid black levels) is broadened by the presence of a continuum of states corresponding to emission or absorption of an LA phonon (grey shading), dressing the optical transition with vibronic bands. 
In the simplest case (Fig. \ref{fig:PSBSpectraRabi}(c)), a photon in the driving field coherently (Rayleigh) scatters directly from the single exciton transition. 
However, in the presence of the phonon environment, the dressing of the optical transition results in non-zero overlaps between vibronic states in the ground and excited state manifolds, such that a scattering event can end in a different vibrational state within the ground-state manifold (Fig. \ref{fig:PSBSpectraRabi}(d)). This corresponds to inelastic Stokes (anti-Stokes) scattering of a lower (higher) energy photon accompanied by the emission (absorption) of an LA phonon, leading to the emergence of a PSB. 
At low bath temperatures, phonon absorption is suppressed, resulting in the characteristic asymmetry of the PSB. 
From Eqs. \ref{eq:G1Opt} and \ref{eq:G1SB}, the branching ratio between phonon-mediated inelastic and elastic  scattering is determined solely by the constant $B^2$. 
Outside the Mollow triplet regime, the coherent ($\mathcal{S} \ll 1$) and incoherent  ($\mathcal{S} \gtrsim 1$) resonant scattering spectra of a SSE thus differ only in the width of the ZPL. 
As such, whilst coherent scattering is often cited as a route to highly coherent single photons, it cannot negate the PSB. 

 \begin{figure*}
    \centering
    \includegraphics{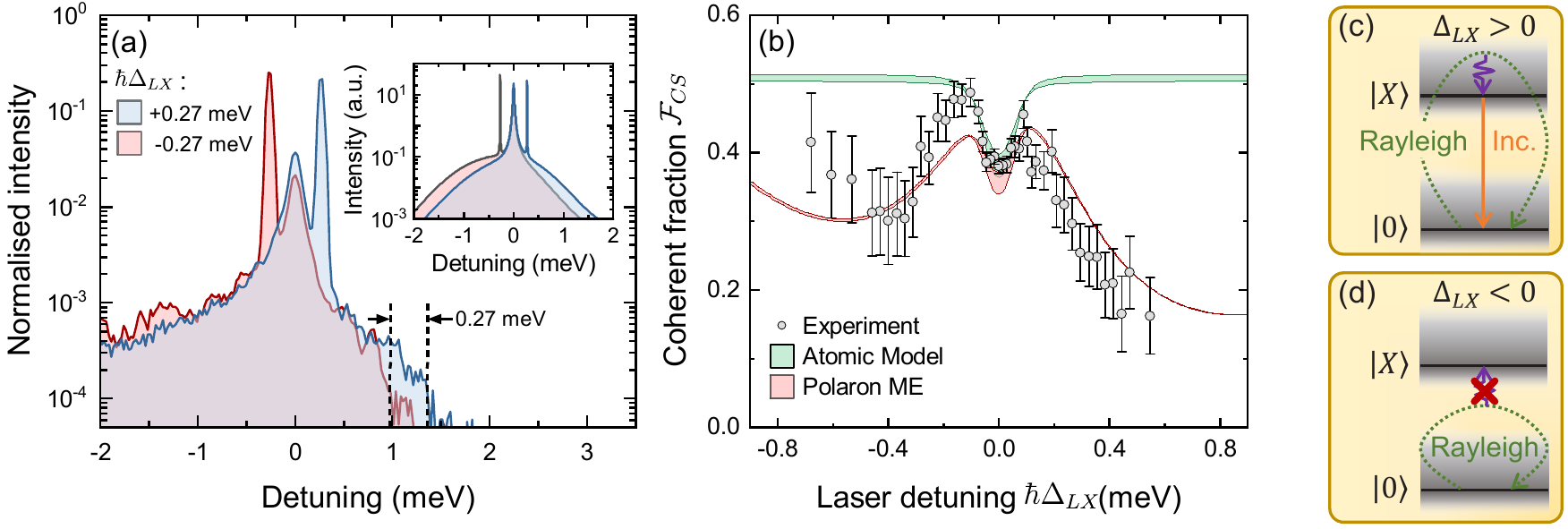}
    \caption{Phonon influences in detuned ($\Delta_{LX} \neq 0$) coherent scattering. (a) Semi-log spectra (normalised by integrated intensity) for $\Delta_{LX}=\pm 0.27~\mathrm{meV}$ (blue/red) at constant $\hbar \Omega = 5.7~\upmu \mathrm{eV}$. 
    \emph{Inset:} Theoretical spectrum.  (b) $\mathcal{F}_{CS}$ vs. $\Delta_{LX}$ at constant $\hbar \Omega=25.6~\upmu \mathrm{eV}$: grey circles - experimental $\mathcal{F}_{CS}$ extracted as in Fig. \ref{fig:PhononG1}(b); red lines - polaron master equation; green lines - atomic model, both models include additional pure dephasing and spectral wandering \cite{Sup} and have upper and lower bounds from uncertainty in $\Omega$. (c) For $\Delta_{LX}>0$, emission of an LA phonon can populate $\ket{X}$, allowing incoherent relaxation. (d) For $\Delta_{LX}<0$, populating $\ket{X}$ requires LA phonon absorption which is weak at $T=4.2~\mathrm{K}$.}
    \label{fig:detuning}
\end{figure*}

To gain further insight into phonon interactions in the scattering picture, the effect of detuning the laser from the emitter is now considered. Fig. \ref{fig:detuning}(a) shows semi-log plots of spectra taken at constant $\Omega$ with laser detuning $\hbar\Delta_{LX}=\pm 0.27~\mathrm{meV}$. 
The coherent peaks at $\hbar \Delta_{LX}$ are separated from the ZPL and dominate the spectrum. For positive detuning (blue spectrum), it is immediately noticeable that the high-energy edge of the sideband is shifted by $\sim \hbar \Delta_{LX}$. The origins of this behaviour can be seen in Eq.~\ref{eq:G1SB}, where the product between $g^{(1)}_\mathrm{opt}(\tau)$ and $(\mathcal{G}(\tau)-B^2)$ in the time-domain implies a convolution in frequency between the purely optical spectrum and the frequency-space phonon correlation function. 
As such, all optical features in $S_\mathrm{opt}$ have an associated PSB; the coherent peak (and thus its PSB) shifts with $\Delta_{LX}$. 
Theoretically (Fig. \ref{fig:detuning}(a) inset), the low-energy edge of the sideband would also be expected to shift for negative detuning (red spectrum); experimentally this is obscured by weak incoherent backgrounds owing to the low count-rate at large $\Delta_{LX}$. 
The total PSB fraction is still governed by $B^2$: since sideband processes arise from phonon dressing of the optical transition, they apply equally to both coherent and incoherent peaks, irrespective of $\omega_L$.

Further deviations from the conventional atomic picture can be seen in the balance of coherent and incoherent scattering when driving off-resonance. 
Compared to the experiment, both the atomic and polaron theories significantly over-estimate the coherent fraction away from resonance (details in SM~\cite{Sup}). 
We attribute this to the Lorentzian reduction in QD absorption with $\vert\Delta_{LX}\vert$, allowing laser light to instead be absorbed in the doped bulk material \cite{doi:10.1063/1.321330}, leading to charge noise. 
To capture the associated pure dephasing in both the atomic and polaron models, we include a Lorentzian detuning-dependent dephasing rate $\gamma(\Delta_{LX})$ with $\gamma(0)=0$, $\gamma(\Delta_{LX} \to \infty)=\gamma_{max}$ and width fixed to the QD linewidth~\cite{Sup}. 
By fitting the polaron theory to the data we find $\gamma_\mathrm{max} = 21\pm0.1~\upmu$eV. 
Spectral wandering is then accounted for by convolving with a Gaussian noise function with 
width deduced from the incoherent peak observed in detuned spectra (Fig. \ref{fig:detuning}(a)) \cite{Sup}.

In Fig. \ref{fig:detuning}(b), upper and lower bounds (from uncertainty in $\Omega$) of the atomic (green curves) and polaron (red curves) models
are plotted. Experimental values of $\mathcal{F}_{CS}$ (grey circles) are evaluated as in Fig. \ref{fig:PhononG1}(b). In stark contrast to the atomic theory, where Eq. \ref{eq:CoherentFracRabi} predicts $\mathcal{F}_{CS}$ will only ever increase with $\vert\Delta_{LX}\vert$, 
the measured data only increases close to resonant driving where EID~\cite{PhysRevLett.104.017402,Ulhaq:13} is small. 
For $\vert \Delta_{LX}\vert$ between 0.1 and 0.4~meV, this EID becomes significant and the coherent fraction decreases with a noticeable asymmetry, as predicted by the polaron model. 
This asymmetry originates from the phonon-dressing of the optical transition: 
when driving above resonance ($\Delta_{LX}>0$) as in Fig. \ref{fig:detuning}(c), $\ket{X}$ can be populated through the emission of an LA phonon~\cite{PhysRevLett.110.147401,Hughes_2013,PhysRevLett.114.137401} (purple arrow), increasing the probability of incoherent scattering (orange arrow).
When $\Delta_{LX}<0$ (Fig. \ref{fig:detuning}(d)), populating $\ket{X}$ is inhibited at $T=4.2~\mathrm{K}$ as it requires phonon absorption~\cite{PhysRevB.93.161407,Brash:16}, inhibiting incoherent scattering. For $\Delta_{LX}<-0.5~\mathrm{meV}$, the probability of phonon absorption becomes sufficiently low that $\mathcal{F}_{CS}$ begins to increase again towards the limiting atomic case.
This behaviour deviates strongly from the atomic model and requires careful consideration for schemes involving detuned coherent scattering, such as generating single~\cite{PhysRevLett.111.237403,Sweeney2014} or entangled~\cite{PhysRevA.96.062329,PhysRevA.98.022318} photons. 

In conclusion, we have shown that a fixed fraction of light scattered from a solid-state emitter is always lost through a phonon sideband, irrespective of excitation conditions such as Rabi frequency or detuning. 
We have also demonstrated that the detuning dependence of the coherent fraction is strongly modified by the presence of phonon coupling, contradicting the atomic prediction that the coherent fraction will increase monotonically with detuning. 
Both processes can be intuitively understood by considering phonon-dressing of the optical transition of the QD. 
Taken together, these results illustrate the importance of employing an appropriate model of phonon coupling rather than assuming atom-like physics when driving weakly or off-resonance. 
For example, treating phonons in a crude pure-dephasing approximation (e.g.  Eq.~\ref{eq:CoherentFracRabi}), suggests they may be suppressed simply by increasing the Purcell factor. This is directly contradicted by the clear separation of phonon and radiative timescales in Fig. \ref{fig:PhononG1}(b), with the phonon sideband persisting despite a large Purcell enhancement.
The methods developed here can be used to optimise quantum information protocols such as spin-photon entanglement schemes for realistic solid-state emitters. 

This work was funded by the EPSRC (UK) EP/N031776/1, A.N. is supported by the EPSRC (UK) EP/N008154/1 and J.I.S. acknowledges support from the Royal Commission for the Exhibition of 1851. \emph{Note:} After the completion of our experiments, we became aware of related results \cite{GerardotarXiv}. We thank B.D. Gerardot for bringing these to our attention.

\bibliography{Bibliography}

\end{document}


\title{Supplemental Material: Light Scattering from Solid-State Quantum Emitters: Beyond the Atomic Picture}

\author{Alistair J. Brash}
\email[Email: ]{a.brash@sheffield.ac.uk}
\affiliation{Department of Physics and Astronomy, University of Sheffield, Sheffield, S3 7RH, United Kingdom}
\author{Jake Iles-Smith}
\email[Email: ]{Jakeilessmith@gmail.com}
\affiliation{Department of Physics and Astronomy, University of Sheffield, Sheffield, S3 7RH, United Kingdom}
\affiliation{School of Physics and Astronomy, The University of Manchester, Oxford Road, Manchester M13 9PL, UK}
\author{Catherine L. Phillips}
\affiliation{Department of Physics and Astronomy, University of Sheffield, Sheffield, S3 7RH, United Kingdom}
\author{Dara P. S. McCutcheon}
\affiliation{Quantum Engineering Technology Labs, H. H. Wills Physics Laboratory and Department of Electrical and Electronic Engineering, University of Bristol, Bristol BS8 1FD, UK}
\author{John O'Hara}
\affiliation{Department of Physics and Astronomy, University of Sheffield, Sheffield, S3 7RH, United Kingdom}
\author{Edmund Clarke}
\affiliation{EPSRC National Epitaxy Facility, Department of Electronic and Electrical Engineering, University of Sheffield, Sheffield, UK}
\author{Benjamin Royall}
\affiliation{Department of Physics and Astronomy, University of Sheffield, Sheffield, S3 7RH, United Kingdom}
\author{Jesper M\o{}rk}
\affiliation{Department of Photonics Engineering, DTU Fotonik, Technical University of Denmark, Building 343, 2800 Kongens Lyngby, Denmark}
\author{Maurice S. Skolnick}
\affiliation{Department of Physics and Astronomy, University of Sheffield, Sheffield, S3 7RH, United Kingdom}
\author{A. Mark Fox}
\affiliation{Department of Physics and Astronomy, University of Sheffield, Sheffield, S3 7RH, United Kingdom}
\author{Ahsan Nazir}
\affiliation{School of Physics and Astronomy, The University of Manchester, Oxford Road, Manchester M13 9PL, UK}

\maketitle


\section{Theoretical Background and Derivations}
We model a quantum dot (QD) as a two level system (TLS) with ground state $\ket{0}$ and a single exciton state $\ket{X}$, with splitting $\omega_X$ ($\hbar=1$). 
The QD is driven by a continuous wave laser with a frequency $\omega_L$ and Rabi coupling $\Omega$. 
The QD couples to both a vibrational environment and a low-Q optical cavity, which is characterised by the Hamiltonian~\cite{Nazir_Review_2016}:
\begin{equation}
H(t) = \omega_X\ket{X}\!\bra{X} + \Omega\cos(\omega_L t)\sigma_x + \ket{X}\!\bra{X}\sum_k g_k\left(b_k^\dagger + b_k\right) + \sigma_x\sum_l \left( h^\ast_la_l^\dagger +h_l  a_l\right) + \sum_l \omega_l a_l^\dagger a_l  + \sum_k\nu_k b^\dagger_k b_k,
\end{equation} 
where we have defined $a_l^\dagger$ as the creation operator for a photon with energy $\omega_l$, and $b_k^\dagger$ as the creation operator for a phonon with energy $\nu_k$.
We have also introduced the system operators $\sigma_x = \sigma^\dagger + \sigma$ and $\sigma = \ket{0}\!\bra{X}$.
The coupling to the vibrational and electromagnetic environments are characterised by their respective spectral densities: for the vibrational environment this takes the standard form $J(\nu)  = \sum_k \vert g_k\vert^2 \delta(\nu-\nu_k) = \alpha\nu^3\exp(-\nu^2/\nu_c^2)$, where $\alpha$ is the deformation potential coupling strength, and $\nu_c$ is the phonon cut-off frequency~\cite{Nazir_Review_2016}; the coupling to the cavity mode is described by a Lorentzian spectral density $\mathcal{J}_C(\omega) = \sum_l\vert h_l\vert^2\delta(\omega-\omega_l) = \pi^{-1}2 g^2\kappa\left[(\omega-\omega_0)^2 + (\kappa/2)^2\right]^{-1}$, where $g$ is the light-matter coupling strength, $\kappa$ is the cavity width, and $\omega_0$ is the cavity resonance. 
Such a treatment of an optical cavity is valid when the cavity loss is much larger than the light-matter coupling strength~\cite{RoyChoudry15,denning2018cavity}.

We can simplify the above equation by making the rotating-wave approximation and moving to a frame rotating with respect to the laser frequency $\omega_L$, yielding the Hamiltonian:
\begin{equation}
H = \Delta\ket{X}\!\bra{X} +\frac{\Omega}{2}\sigma_x + \ket{X}\!\bra{X}\sum_k g_k (b_k^\dagger + b_k) + \sum_l \left(h^\ast_l\sigma a_l^\dagger e^{i\omega_L t} +h_l  \sigma^\dagger a_l e^{-i\omega_L t}\right)
+ \sum_l \omega_l a_l^\dagger a_l  + \sum_k\nu_k b^\dagger_k b_k,
  \end{equation}
where $\Delta_\mathrm{LX} = \omega_X - \omega_L$ is the detuning between the driving field and the exciton transition. 
This Hamiltonian forms the starting point of our analysis of the dynamical and optical properties of the QD.

\subsection{Polaron theory for a driven emitter in the Purcell regime}
In order to account for the strong coupling to the vibrational environment, we apply a polaron transformation to the global Hamiltonian, i.e. the unitary transformation $\mathcal{U}_P = \ket{X}\!\bra{X}\otimes B_+  + \ket{0}\!\bra{0}$, where $B_\pm = \exp\left(\pm\sum_k \nu_k^{-1} g_k(b_k^\dagger - b_k)\right)$ are displacement operators of the phonon environment~\cite{McCutcheon_2010,Nazir_Review_2016,PhysRevB.95.201305}. This transformation dresses the excitonic states with vibrational modes of the phonon environment.
In the Polaron frame, the Hamiltonian may be written as $H_P= H_0  + H^\mathrm{PH}_\mathrm{I}+ H^\mathrm{EM}_\mathrm{I}$, where  
\begin{equation}
\begin{split}
H_0& = \tilde\Delta\ket{X}\!\bra{X} + \frac{\Omega_R}{2}\sigma_x + \sum_l \omega_l a_l^\dagger a_l  + \sum_k\nu_k b^\dagger_k b_k,\\
H^\mathrm{PH}_\mathrm{I}& = \frac{\Omega}{2}\left(\sigma_x B_x + \sigma_y B_y\right),\hspace{0.5cm}\mathrm{and}\hspace{0.5cm} H^\mathrm{EM}_\mathrm{I}  = \sum_l h_l ^\ast \sigma B_+ a_l^\dagger e^{i\omega_L t}+ \mathrm{h.c.},
\end{split}
\end{equation}
where we have introduced the phonon operators $B_x = (B_+ + B_- - 2B)/2$, $B_y = i(B_+ - B_-)/2$, and the Frank-Condon factor of the phonon environment $B = \tr_B(B_\pm\rho_B) = \exp(-(1/2)\sum_k \nu_k^{-2}\vert{}g_k\vert^2\coth(\nu_k/k_BT))$, with the Gibbs state of the phonon environment in the polaron frame given by $\rho_B = \exp(-\sum_k\nu_k b_k^\dagger b_k/k_BT)/\tr\left(\exp(-\sum_k\nu_k b_k^\dagger b_k/k_BT)\right)$. 
Notice that the polaron transformation has dressed the operators in the light-matter coupling Hamiltonian, and renormalised the system parameters, such that $\Omega_R= \Omega B$ and $\tilde{\Delta} = \Delta - \sum_k \nu_k^{-1} g^2_k$.

To describe the dynamics of the QD, we now proceed to derive a 2$^\textrm{nd}$-order Born Markov master equation~\cite{breuer2002theory}, perturbatively eliminating both the electromagnetic and vibrational environments. 
We can use the fact that only terms quadratic in field operators are non-zero when traced with a Gibbs state, so that there will be no cross-terms between the vibrational and electromagnetic dissipators, such that the master equation can be written in two parts:
\begin{equation}
\frac{\partial\rho(t)}{\partial t} = -i\left[\tilde\Delta\ket{X}\!\bra{X} + \frac{\Omega}{2}\sigma_x,\rho(t)\right]+ \mathcal{K}_\mathrm{PH}[\rho(t)] + \mathcal{K}_\mathrm{EM}[\rho(t)]
\end{equation}
where $\rho(t)$ is the reduced state of the QD, $\mathcal{K}_\mathrm{EM}$ is the dissipator for the electromagnetic environment and similarly $\mathcal{K}_\mathrm{PH}$ describes the dissipation due to the vibrational environment. 
In the following we outline how these dissipators are derived.

\subsubsection{Deriving the phonon dissipator}
We start by considering only the coupling to the vibrational modes. 
Moving into the interaction picture with respect to the Hamiltonian $H_0$, the interaction term becomes:
\begin{equation}
H_\mathrm{I}^\mathrm{PH}(t) = \frac{\Omega}{2} (\sigma_x(t) B_x(t) + \sigma_y(t) B_y(t)),
\end{equation}
where the interaction picture system operators are given by:
\begin{equation}\begin{split}
\sigma_x(t)& = \eta^{-2}\left[\tilde\Delta\Omega_R(1-\cos(\eta t))\sigma_z + (\Omega_R^2 + \tilde\Delta^2\cos(\eta t))\sigma_x + \tilde\Delta \eta\sin(\eta t)\sigma_y
\right],\\
\sigma_y(t)&= \eta^{-1}\left(\Omega_R\sin(\eta t)\sigma_z + \eta\cos(\eta t)\sigma_y - \tilde\Delta\sin(\eta t)\sigma_x\right),
\end{split}\end{equation}
and we have defined the generalised Rabi frequency $\eta = \sqrt{\tilde\Delta^2 + \Omega_R^2}$.
In the Schr\"odinger picture and making the Born-Markov approximation in the polaron frame~\cite{breuer2002theory}, the dissipator describing the electron-phonon interaction is given by: 
\begin{equation}
\mathcal{K}_\mathrm{PH} [\rho(t)]  = -\frac{\Omega^2}{4}\int\limits_0^\infty \left([\sigma_x,\sigma_x(-\tau)\rho(t)]\Lambda_{xx}(\tau) + [\sigma_y,\sigma_y(-\tau)\rho(t)]\Lambda_{yy}(\tau) + \operatorname{h.c.} \right)d\tau,
\end{equation}
where we have and introduced the phonon bath correlation functions as 
$
\Lambda_{xx}(\tau) = \langle B_x(\tau)B_x\rangle = B^2(e^{\varphi(\tau)} + e^{-\varphi(\tau)} - 2)
$, 
$\Lambda_{yy}(\tau) =  \langle B_y(\tau)B_y\rangle  = B^2(e^{\varphi(\tau)}-e^{-\varphi(\tau)})$. Taking the continuum limit over the phonon modes, the polaron propagator is given by $\varphi(\tau) = \int_0^\infty d\nu \nu^{-2}J(\nu)(\coth(\beta\nu/2)\cos(\nu\tau)- i\sin(\nu\tau)$.
We can simplify the form of the master equation by evaluating the integrals over $\tau$, leading to:
\begin{equation}
\mathcal{K}_\mathrm{PH}[\rho(t)] = -\frac{\Omega^2}{4} \left([\sigma_x, \chi_x\rho_s(t)] + [\sigma_y, \chi_y\rho_s(t)] + \operatorname{h.c.}\right).
\end{equation}
where we have introduced the rate operators:
\begin{align}
\chi_x &= \int\limits_0^\infty \sigma_x(-\tau)\Lambda_{xx} (\tau) d\tau = \frac{1}{\eta^2}\left[\Delta\Omega_r(\Gamma^x_0 - \Gamma^x_c) \sigma_z + (\Omega_r^2 \Gamma^x_0+ \Delta^2\Gamma^x_c)\sigma_x + \Delta\eta \Gamma^x_s \sigma_y\right] ,\\
\chi_y & = \int\limits_0^\infty \sigma_y(-\tau)\Lambda_{yy} (\tau) d\tau = \frac{1}{\eta}\left[\Omega_r\Gamma^y_s \sigma_z - \Delta\Gamma^y_s\sigma_x + \eta \Gamma^y_c \sigma_y\right],
\end{align}
and defined the rates $\Gamma^a_0 = \int_0^\infty\Lambda_{aa}(\tau)d\tau$, $\Gamma^a_c = \int_0^\infty\Lambda_{aa}(\tau)\cos(\eta\tau)d\tau$, $\Gamma^a_s = \int_0^\infty \Lambda_{aa}(\tau)\sin(\eta\tau)d\tau$, with $a \in \{x,y\}$.

Since we are operating in the polaron frame, the above master equation is non-perturbative in the electron-phonon coupling strength, capturing strong-coupling and non-Markovian influences in the lab frame, despite having made a Born Markov approximation. This provides us with a simple, intuitive, and computationally straight-forward method for describing exciton dynamics in a regime where a standard weak coupling master equation would break down~\cite{Nazir_Review_2016}.

\subsubsection{Deriving the electromagnetic dissipator}
We now consider the case of the electromagnetic field coupling. 
If we once again move to the interaction picture with respect to the free Hamiltonian $H_0$, then the interaction Hamiltonian may be written as:
\begin{equation}
H^\mathrm{I}_\mathrm{EM}(t) = \sigma(t) B_+(t)\hat{A}^\dagger(t) e^{i\omega_L  t} + \sigma^\dagger(t) B_-(t)\hat{A}(t) e^{i\omega_L  t},
\end{equation}
where $A(t) = \sum_l h_l a_l(t)$.
Using this Hamiltonian, we can derive a second-order master equation such that, in the Sch\"odinger picture, the dissipator may be written as:
\begin{equation}
\mathcal{K}_\mathrm{EM}[\rho(t)] = -\int\limits_0^\infty d\tau\left(\left[\sigma, \sigma^\dagger(-\tau)\rho(t)\right]\langle B_+(\tau) B_-\rangle\langle \hat{A}^\dagger(\tau) \hat{A}\rangle+\left[\sigma^\dagger, \sigma(-\tau)\rho(t)\right]\langle B_-(\tau) B_+\rangle\langle \hat{A}(\tau) \hat{A}^\dagger\rangle + \mathrm{h.c.}\right),
\end{equation}
where we have used the Born approximation to factorise the vibrational and electromagnetic correlation functions.
We take the cavity field to be at zero temperature, allowing for only spontaneous emission processes, such that $\langle \hat{A}^\dagger(t)\hat{A}\rangle = 0$. 
The master equation is then written compactly as
$
\mathcal{K}_{EM}[\rho(t)] = -\left(\left[\sigma^\dagger, \varsigma\rho(t)\right]+ \left[\rho(t)\varsigma^\dagger, \sigma\right]\right)
$
where we have defined the rate operator 
$
\varsigma = \int_0^\infty d\tau \sigma(-\tau)\langle B_+(\tau) B_-\rangle \langle\hat{A}(\tau)\hat{A}^\dagger\rangle
$.
The field correlation function is given by:
\begin{equation}
\langle\hat{A}(\tau)\hat{A}^\dagger\rangle = \int\limits_{-\infty}^\infty \mathcal{J}_C(\omega)e^{i(\omega +\omega_L)\tau}d\omega = g^2 e^{[-i(\omega_c +\omega_L) - \kappa/2]\tau} , 
\end{equation}
where we have extended the lower limit of integration to $-\infty$, which is justified when the peak of the optical spectral density is far from the origin.
The total rate operator becomes~\cite{RoyChoudry15}:
\begin{equation}
\varsigma = g^2 B^2\int_0^\infty d\tau \sigma(-\tau) e^{-[i(\omega_c +\omega_L) +\kappa/2]\tau}   e^{\varphi(\tau)}.
\end{equation}
We can simplify this expression by recognising that the spectral density of the cavity does not vary appreciable over the energy scale relevant for the driving field, $\Omega_R$. This allows us to approximate the interaction picture transformation as $\sigma(t) \approx \sigma e^{-i\Delta t}$, such that the master equation can be written in Lindblad form:
\begin{equation}
\mathcal{K}_{EM}[\rho(t)] =-\frac{iS}{2}\left[\sigma^\dagger\sigma,\rho(t)\right] + \frac{\Gamma_P}{2}\mathcal{L}_{\sigma}[\rho(t)],
\end{equation}
where $\mathcal{L}_{O}[\rho(t)] = 2O\rho(t)O^\dagger - \{O^\dagger O, \rho(t)\}$, $S = \mathrm{Im}(\Gamma)$ and $\Gamma_P = \mathrm{Re}(\Gamma)$, with:
\begin{equation}
{\Gamma}=2 {g^2 B^2}\int\limits_0^\infty  e^{\varphi(t)}e^{-\left[i(\omega_c - \tilde\omega_X) +\kappa/2\right]\tau}d\tau ,
\end{equation}
and $\tilde{\omega}_X = \omega_X - \sum_k \nu_k^{-1}g_k^2$ is the polaron shifted exciton transition.

\subsection{Correlation functions and spectra for a driven QD}
We are interested in understanding the impact that phonon coupling has on the coherence properties of the scattered field.
This can be calculated from the steady-state first-order correlation function:
\begin{equation}
g^{(1)}(\tau) = \lim_{t\rightarrow\infty}\left\langle\hat{E}^\dagger(t + \tau)\hat{E}(t)\right\rangle,
\end{equation}
where we have defined Heisenberg picture electric field operators as $\hat{E}(t) = \sum_l a_l(t)$. 

\subsubsection{Scattering in the presence of a low-Q cavity}
In this section, we shall outline how the $g^{(1)}$ may be written in terms of system operators in the presence of the low-Q cavity mode, and demonstrate that the dominant spectral feature of the cavity in this regime is to filter the QD emission.
This is most easily done by formally relating the field and system operators in the Heisenberg frame~\cite{Iles-Smith2017}.
The Heisenberg equations of motion for the field operators in the polaron frame are given by:
$
\partial_t a_l(t) = -i\omega_l a_l(t) - i h^\ast_l \sigma(t)B_+(t) e^{i\omega_L t}.
$
We can write a formal solution for this equation such that:
\begin{equation}
\hat{a}_l(t) = \hat{a}_l(0)e^{-i\omega_l  t} -ih_l \int\limits_{0}^t\underline{\bm\sigma}(t^\prime)e^{i\omega_L t^\prime}e^{i\omega_l(t^\prime - t)}dt^\prime,
\end{equation} 
where we have defined the polaronic dipole operator $\underline{\bm\sigma} = \sigma B_+$ for convenience. 
The first term in this expression describes free evolution of the field, and can be discarded if we assume the photonic environment is initially in its vacuum state.
Substituting this solution into the expression for the field operators, and taking the continuum limit of the field modes, we obtain:
\begin{equation}
E(t) = \sum_l \hat{a}_l(t)\rightarrow - i\int\limits_{t_0}^t dt^\prime\int\limits_{0}^\infty d\omega h(\omega) e^{i\omega(t^\prime - t)}e^{i\omega_L t^\prime}\underline{\bm\sigma}(t^\prime),
\end{equation}
where we have introduced the continuum coupling function $h(\omega)$. For an optical cavity, this takes a complex Lorentzian form such that~\cite{RoyChoudry15}:
\begin{equation}
h(\omega) = \frac{1}{\sqrt{\pi}}\frac{\sqrt{2g^2\kappa}}{i(\omega-\omega_c) + \kappa/2}.
\end{equation}
The frequency integral can be analytically evaluated by extending the lower integration limit to $-\infty$, yielding:
\begin{equation}
E(t) =-i\sqrt{8\pi g^2\kappa} \int\limits_{0}^t dt^\prime e^{-(i(\omega_c -\omega_L)+ \kappa/2)(t^\prime-t)} \underline{\bm\sigma}(t^\prime) = -i \int\limits_{0}^t dt^\prime f(t^\prime-t) \underline{\bm\sigma}(t^\prime),
\end{equation}
where we see that the field operator is written as a convolution over a time domain filter function $f(t)$ and the polaron dipole operator $\underline{\bm\sigma}(t)$, representing the cavity mode. 

Using the above expression for the field operators, we can formally write the first-order correlation function in terms of QD operators, such that:
\begin{equation}
\begin{split}
g^{(1)}(\tau) =& \lim_{t\rightarrow\infty}\int\limits_{-\infty}^{t+\tau} dt_1\int\limits_{-\infty}^{t} dt_2
f^\ast(t+\tau-t_1) f(t - t_2)\langle\underline{\bm\sigma}^\dagger(t_1)\underline{\bm\sigma}(t_2)\rangle,
\end{split}
\label{g1Definition}
\end{equation}
taking the initial time to be $ t_0=-\infty$.
 In order to simplify this expression, we want to commute the limit through the integrals. %
This can be done through the change of variables $r = 2t + \tau - (t_1+ t_2)$  and $s = t_1 - t_2$, such that $t_1=(1/2)(s-r) +t + \tau/2$, and $t_2 = -(1/2)(s+r) + t + \tau/2$. To find the area element defined by this change of variable, we calculate the Jacobian $\vert\partial_s t_1 \partial_r t_2 - \partial_r t_1 \partial_s t_2\vert = 1/2$, such that $\operatorname{d}\!t_1 \operatorname{d}\!t_2 = (1/2)\operatorname{d}\!s \operatorname{d}\!r$. 
The integration limits are can then found to be $-\infty<s<\infty$ and $\vert s- \tau\vert< r<\infty$.
Put together this yields a correlation function of the form:
\begin{equation}
g^{(1)}(\tau) = 4\pi g^2\kappa\lim_{t\rightarrow\infty} \int\limits_{-\infty}^\infty\! ds\int\limits_{\vert s- \tau\vert}^\infty \!dr e^{i(\omega_c-\omega_L) (s-\tau)}e^{-(\kappa/2) r}\left\langle\underline{\bm\sigma}^\dagger\left(\frac{1}{2}(s-r) +t + \tau/2\right)\underline{\bm\sigma}\left(-\frac{1}{2}(s+r) +t + \tau/2\right)\right\rangle.
\end{equation}
As we see, the $t$-dependence only occurs in the correlation functions, so we can permute the limit through the integrals. This also allows us to use the property of the correlation function that $\lim_{t\rightarrow\infty} \langle\underline{\bm\sigma}^\dagger(x_1 + t) \underline{\bm\sigma}(x_2 + t)\rangle = \lim_{t\rightarrow\infty} \langle\underline{\bm\sigma}^\dagger(x_1 - x_2 + t) \underline{\bm\sigma}( t)\rangle= \langle\underline{\bm\sigma}^\dagger(x_1 - x_2) \underline{\bm\sigma}\rangle_{ss}$, 
such that the $r$ integral can solved analytically, yielding:
$
g^{(1)}(\tau) =  \int_{-\infty}^{\infty}\!ds \tilde{h}(s-\tau)g^{(1)}_0(s),
\label{eq:fullg1}
$
where $g^{(1)}_0(s)=\langle\underline{\bm\sigma}^\dagger(s)\underline{\bm\sigma}(0)\rangle_{ss}$ and $\tilde{h}(x) = 8\pi g^2\exp[i(\omega_c -\omega_L)x-(\kappa/2)\vert x\vert]$. 
Notice that this correlation function is a convolution between the time-domain filter function of the cavity, and the optical first-order correlation function. 
By making use of the convolution theorem, we may immediately write the spectrum emitted from the cavity: 
\begin{equation}
S_C(\omega) = H(\omega)S(\omega),
\end{equation}
where $H(\omega) = 8\pi g^2\kappa /[(\omega - (\omega_L - \omega_c))^2 + (\kappa/2)^2]$ is the frequency-domain cavity filter function and  $S(\omega) = \int_{-\infty}^\infty g^{(1)}_0(\tau)\exp(i\omega\tau)d\tau$ is the spectrum of the emitter.

As a final point, we note that the regression theorem only is defined for $\tau\geq0$. To ensure this is satisfied in the time-domain detected correlation function, we may re-order the limits, such that:
\begin{equation}
g^{(1)}(\tau) =\int\limits_{0}^{\infty}\!ds \tilde{h}(s-\tau)g^{(1)}_0(s)
+\int\limits_{0}^{\infty}\!ds\tilde{h}^\ast(-s-\tau)(g^{(1)}_0(s))^\ast
.
\end{equation}
Similarly, for the spectra we can restrict ourselves to positive times, by writing $S(\omega)= \mathrm{Re}[\int_0^\infty g^{(1)}_0(\tau)\exp(i\omega\tau)d\tau]$.

\subsubsection{Emission in the polaron frame: non-Markovianity and the emergence of the phonon sideband}
As shown in the previous section, the first-order correlation function for the emitter is written in terms of the polaronic dipole operator $\underline{\underline{\underline{\underline{\bm\sigma}}}} = \sigma B_+$, that is, the correlation function is $g^{(1)}_0(\tau) = \lim_{t\rightarrow\infty}\langle B_-(t+\tau)\sigma^\dagger(t+\tau) B_+(t)\sigma(t)\rangle$.
In general, $B_\pm$ are many-body displacement operators that do not commute with system operators. 
However, in the regime that the polaron master equation is valid at second-order, we can factor this correlation function in the Born approximation, such that $g^{(1)}_0(\tau) \approx \mathcal{G}(\tau)g^{(1)}_\mathrm{opt}(\tau)$.
There are two terms in this expression, the correlation function for the scattered field $g^{(1)}_{opt}(\tau) = \langle\sigma^\dagger(\tau)\sigma\rangle_{ss}$ describing the purely electronic transitions, and the phonon correlation function $\mathcal{G}(\tau) = \langle B_-(\tau)B_+\rangle = B^2\exp(\varphi(\tau))$ which accounts for the relaxation of the phonon environment.

This factorisation allows us to split the emission spectrum into two components $S(\omega) =S_\mathrm{opt}(\omega) + S_\mathrm{SB}(\omega)$, where $S_\mathrm{opt}(\omega) =\mathrm{Re}[B^2\int_0^\infty g^{(1)}_\mathrm{opt}(\tau)e^{i\omega\tau}d\tau]$ corresponds to purely optical transitions, and $S_\mathrm{SB}(\omega) =\mathrm{Re}[\int_0^\infty (\mathcal{G}(\tau)-B^2)g^{(1)}_\mathrm{opt}(\tau)e^{i\omega\tau}d\tau]$ is attributed to non-Markovian phonon relaxation during the emission process, and leads to the emergence of a phonon sideband.

\subsubsection{Coherent and incoherent scattering in the presence of non-Markovian phonon processes}
We are interested in understanding the coherence properties of the scattered field, and the impact phonon processes have on it. To investigate this, we divide the correlation function into a coherent and incoherent contribution, that is $g^{(1)}_\mathrm{opt} (\tau)= g_\mathrm{inc}(\tau) +g_\mathrm{coh}$, where $g_\mathrm{coh} = \lim_{\tau\rightarrow\infty}g^{(1)}_\mathrm{opt}(\tau)$ and $g_\mathrm{inc}(\tau) =g^{(1)}_\mathrm{opt} (\tau)-g_\mathrm{coh}$. 
This allows us to further divide the emission spectrum into two components $S_\mathrm{opt}(\omega) = S_\mathrm{inc}(\omega)+S_\mathrm{coh}(\omega)$, where $S_\mathrm{inc}(\omega)=\mathrm{Re}[B^2\int_0^\infty g^{(1)}_\mathrm{inc}(\tau)e^{i\omega\tau}d\tau]$
and $S_\mathrm{coh}(\omega)=\pi B^2g_\mathrm{coh}\delta(\omega)$ with $\delta(\omega)$ the Dirac $\delta$-function.
We want to find the fraction of light scattered coherently, which can be done by finding the integrated powers. 
The total power is found as $\mathcal{P}_\mathrm{tot} = \int_{-\infty}^\infty H(\omega)S(\omega)d\omega$, and the coherent power is $\mathcal{P}_\mathrm{coh} = B^2H(0)g_\mathrm{coh}$. The fraction of coherently scattered light is then given by the ratio of the powers $\mathcal{F}_\mathrm{CS} = \mathcal{P}_\mathrm{coh}/\mathcal{P}_\mathrm{tot}$.

\section{Experimental Details}

In this section we outline some additional details of the measurement setup and methods.

\subsection{Interferometry - first-order correlation function $g^{(1)}(\tau)$}

\subsubsection{Measurement of fringe contrast}

\begin{figure}
    \centering
    \includegraphics{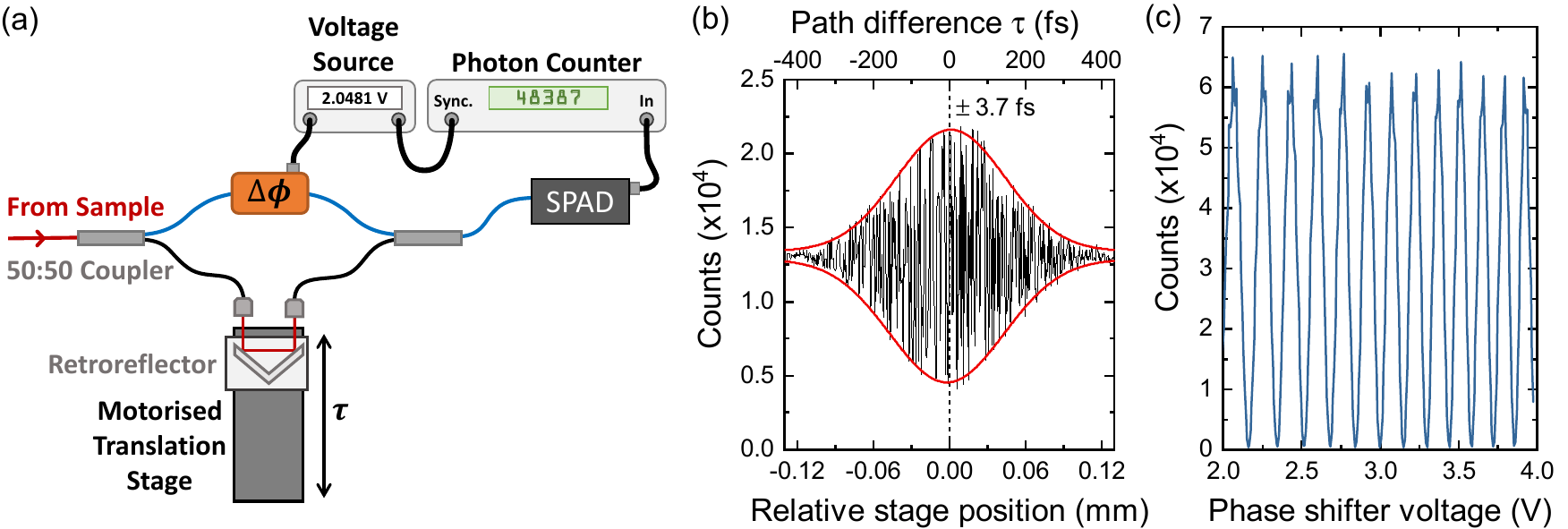}
    \caption{Details of the $g^{(1)}(\tau)$ measurement. (a) Schematic of the fiber Mach-Zehnder interferometer. A fiber phase shifter provides a phase shift ($\Delta \phi$) to measure fringe contrast whilst a motorised translation stage controls the path difference ($\tau$) between the two arms. The output intensity is detected by a single photon avalanche diode (SPAD) and recorded by a gated photon counter. (b) Calibration of the path difference ($\tau$) between the interferometer arms. $\tau=0$ is found from Gaussian fits (red lines) to the envelope of the injected ultrafast laser pulse. (c) Example plot of fringes recorded by the photon counter as a function of $\Delta \phi$ by changing the voltage applied to the fiber phase shifter.}
    \label{fig:Supg1t}
\end{figure}

A schematic of the setup used for the first-order correlation measurements is presented in Fig. \ref{fig:Supg1t}(a). After cross-polarised rejection of the scattered laser, the signal is passed to a fiber Mach-Zehnder interferometer, containing a fiber phase shifter ($\Delta \phi$) in one arm and a motorized translation stage in the other. The intensities of the two arms are balanced with a variable fiber attenuator and their polarizations matched (typical extinction $\sim 1000:1$) using a pair of fiber paddles on each arm. The interferometer is placed in a thermally stable environment to minimise any phase drifts.

The stage position at which the two arms are of equal length ($\tau = 0$) is found by injecting ultrafast pulses from a Ti:Sapphire laser emitting at the same wavelength as the QD into the interferometer, scanning the translation stage position and finding the maximum of the envelope (red lines - Gaussian fit) of the interference fringes as illustrated in Fig. \ref{fig:Supg1t}(b). This calibration is made with the fiber phase shifter in the centre of its scan range.

To measure the fringe contrast for a given $\tau$, $\Delta \phi$ is scanned by applying a voltage sweep (step size $\sim 0.1\pi$, sweep rate $\sim 2\pi/3~\mathrm{s}^{-1}$) to the fiber phase shifter. A synchronisation signal from the voltage source is passed to a gated photon counter which records the count-rate of a single photon avalanche detector (SPAD) connected to an output port of the interferometer. An example of the trace recorded by the photon counter is shown in Fig. \ref{fig:Supg1t}(c). We perform a generalised peak-finding routine to find the intensity at the local maximas ($I_{max}$) and minimas ($I_{min}$) of the data and then evaluate the fringe contrast ($v$) according to:
\begin{equation}
    v = \frac{I_{max}-I_{min}}{I_{max}+I_{min}}.
\end{equation}
The maximum resolvable contrast (defined as $1-\epsilon$) can be checked by injecting a long coherence time single mode laser. For the interferometer used here it is limited to $(1 - \epsilon) = 0.98$ by the imperfect mode overlap of the second 50:50 fiber coupler. For low count-rates (e.g. large $\Delta_{LX}$), shot-noise begins to further limit the maximimum measurable contrast and also causes the associated uncertainty to increase.

\subsubsection{Relationship between fringe contrast and $g^{(1)}(\tau)$}

To see how this measured fringe contrast visibility relates to the first-order correlation function of the scattered field, 
we consider the mode transformation that maps the scattered electric field operator $E_{\mathrm{in}}(t)$ to that 
detected at the SPAD, which we label $E_{\mathrm{out}}(t)$. Neglecting retardation effects, 
the Mach-Zehnder described above amounts to the transformation
\begin{equation}
E_{\mathrm{out}}(t)=\frac{1}{2}\big(E_{\mathrm{in}} (t+\tau_{\Delta\phi})+E_{\mathrm{in}}(t+\tau)\big),
\end{equation}
where $\tau_{\Delta\phi}$ is the time delay corresponding to the phase difference $\Delta\phi$. 
The intensity measured at the SPAD in the long time limit is then a function of $\tau_{\Delta\phi}$ and $\tau$, and is given by 
\begin{align}
I(\Delta\phi,\tau)=\lim_{t\to\infty} \langle E^{\dagger}_{\mathrm{out}}(t) E_{\mathrm{out}}(t)\rangle 
&= \frac{1}{2}g^{(1)}(0)+\frac{1}{2}\mathrm{Re}\big[g^{(1)}(\tau-\tau_{\Delta\phi})\mathrm{e}^{i(\tau-\tau_{\Delta\phi})\omega_L}\big],
\end{align}
where $g^{(1)}(\tau)$ is the rotating frame steady-state first-order correlation function of the emitted field as before, c.f. Eq.~\ref{g1Definition}, 
and the exponential factor appears in order to convert this into a non-rotating frame measured quantity. 
If we now write $g^{(1)}(\tau)=|g^{(1)}(\tau)|\exp[i \theta(\tau)]$, where $\theta(\tau)$ is some real function 
capturing the phase of the correlation function, we find 
\begin{align}
I(\Delta\phi,\tau) = \frac{1}{2}g^{(1)}(0)+\frac{1}{2}\cos[\theta(\tau-\tau_{\Delta\phi})+\tau\omega_L-\Delta\phi]|g^{(1)}(\tau-\tau_{\Delta\phi})|,
\end{align}
where $\Delta\phi=\tau_{\Delta\phi}\omega_L$. For fixed $\tau$ and varying $\Delta\phi$, we then see that the measured intensity 
has maxima and minima given by
\begin{equation}
I(\Delta\phi,\tau) = \frac{1}{2}g^{(1)}(0)\pm\frac{1}{2}|g^{(1)}(\tau-\tau_{\Delta\phi})|,
\end{equation}
which correspond to the peaks and troughs seen in Fig.~\ref{fig:Supg1t}(c).

In our measurements, the values of $I_{max}$ and $I_{min}$ used to find the visibility 
for a fixed delay $\tau$ are 
found by varying the phase $\Delta\phi$ from approximately $-15\pi$ to $+15\pi$ and 
taking the average value of the maxima and minima. This corresponds to 
phase delays of approximately $\tau_{\Delta\phi}=\Delta\phi/\omega_L\approx 0.04~\mathrm{ps}$. 
As such the extracted values can be written 
\begin{equation}
I_{max}=\frac{1}{2}\Big(g^{(1)}(0)+|g^{(1)}(\tau)|\Big)\qquad \mathrm{and}\qquad
I_{min}=\frac{1}{2}\Big(g^{(1)}(0)-|g^{(1)}(\tau)|\Big)
\end{equation}
where the correlation function $|g^{(1)}(\tau)|$ written above should be thought of as a 
time-averaged coarse grained value with resolution $\sim 0.1~\mathrm{ps}$. 
Finally, we see that the measured 
fringe contrast visibility is then related to the emitted field first-order correlation function through
\begin{equation}
v(\tau)=(1-\epsilon)\frac{|g^{(1)}(\tau)|}{g^{(1)}(0)},
\end{equation}
where $(1-\epsilon)$ is the maximum resolvable fringe contrast of the interferometer as discussed in the previous section. Accounting for $(1-\epsilon)$, it is thus shown that $v(\tau)$ corresponds to the absolute value of the normalised coarse grained first-order correlation function.

\subsection{Spectroscopy}

In Fig. 2(a) of the main text, the resonance fluorescence spectrum of the QD was presented. Here we outline the procedure used to extract the coherent and incoherent contributions for a typical spectrum. 
The ZPL is fitted with a Voigt profile (yellow line) which incorporates the Gaussian spectrometer instrument response. The PSB contribution ($S_{SB}(\omega)$) is well-approximated by a simple Gaussian function (red curve). The PSB fraction $\mathcal{F}_{PSB}$ is then evaluated from the areas ($A$) of these fits according to:
\begin{equation}
    \mathcal{F}_{PSB} = \frac{A_{PSB}}{A_{PSB}+A_{ZPL}}.
\end{equation}

Performing high resolution spectroscopy of the ZPL with the Fabry-Perot interferometer (FPI) results in the inset to Fig. 2(a). As the signal is first filtered (FWHM $96~\upmu \mathrm{eV}$) by the spectrometer which is centered on the ZPL (see Fig. 1(a) of main text), the PSB contribution is removed from the high resolution measurement.
Within the high-resolution spectrum of the ZPL, the broad component (blue) is the Lorentzian spectrum of the incoherent resonance fluorescence, with a linewidth governed by the transition coherence time $T_2$. The narrow (green) component is attributed to coherent scattering; the linewidth observed in this measurement is limited by the Gaussian IRF ($\sim 0.5~\upmu\mathrm{eV}$) caused by drifts of the FPI during the measurement.
We note that the natural linewidth of the incoherent peak ($\sim 20~\upmu\mathrm{eV}$ owing to the short $T_2$) is comparable to the free-spectral range (FSR) of the FPI; to account for this we fit the sum of a Lorentz and a Gaussian function to the 3 peaks that occur in a scan over 3 FSRs so that the central fit accurately accounts for contributions from adjacent FSRs.
After fitting, the coherent ($\mathcal{F}_{CS}$) and incoherent ($\mathcal{F}_{INC}$) fractions of the emission may then be calculated from the fitted areas ($A$) of both the low and high resolution spectra according to:
\begin{equation}
    \mathcal{F}_{CS} = \frac{A_{CS}}{A_{CS}+A_{INC}} \frac{A_{ZPL}}{A_{PSB}+A_{ZPL}},
\end{equation}
and:
\begin{equation}
    \mathcal{F}_{INC} = \frac{A_{INC}}{A_{CS}+A_{INC}} \frac{A_{ZPL}}{A_{PSB}+A_{ZPL}},
\end{equation}
respectively, producing the values used in Fig. 2(b) of the main text.

\section{Extracting phonon parameters from experimental data}

In this section we briefly outline the fitting procedure used to characterise the electron-phonon interaction, and interpret experimental data.

\subsection{Phonon dynamics in the time-domain}
Now that we have expressions for the first-order correlation function, we can approach fitting the $g^{(1)}$ extracted from the experimental data. 
The $g^{(1)}$ measured at low power, shown in Fig. 1(b) of the manuscript, can be partitioned into three separate components: below $15$~ps we see a rapid decay of coherence, which we attribute to non-Markovian relaxation of the phonon environment; after this there is a slower radiative decay and ultimately a plateau at longer times, these features are attributed to incoherent and coherent scattering respectively.

To extract the phonon parameters, we therefore need only to focus on the dynamics up to 15~ps. 
In this situation, we can drastically simplify the expression for the correlation function, by replacing the zero-phonon line contribution with its initial value, that is the unfiltered correlation function becomes $g_0^{(1)}(\tau<15\mathrm{~ps}) = \mathcal{G}(\tau)g^{(1)}_\mathrm{opt}(\tau)\approx\langle\sigma^\dagger\sigma\rangle_{ss}\mathcal{G}(\tau)$.
Normalising this filtered correlation function to its value at zero time delay and restricting ourselves to positive time delays, we obtain the fitting function:
\begin{equation}
g^{(1)}_{F}(\tau, \alpha, \nu_c) = \left[\operatorname{Re}\left(\int\limits_0^\infty \tilde{h}(s)\mathcal{G}(s) ds\right)\right]^{-1}\int\limits_0^\infty ds
\left[
\tilde{h}(s-\tau)\mathcal{G}(s) + \tilde{h}(-s-\tau)\mathcal{G}^\ast(s)\right].
\end{equation}
We are now left with an expression containing two fitting parameters: the electron phonon coupling strength $\alpha$, and the cut-off frequency $\nu_c$.
Through a simple root mean squared fitting, we find $\alpha = 0.0447$~ps$^2$ and  $\nu_c = 1.28$~ps$^{-1}$.

\subsection{Instrument response in the weak-driving regime}
When driving below saturation, one expects the majority of scattered photons to have linewidths matching the driving laser. This linewidth cannot be resolved in the current set-up due to a finite instrument response of the spectrometer, meaning that the measured spectrum must be convolved with a detector response function, which is assumed to be Gaussian.

We can do so in the time-domain by multiplying the detected correlation function by the instrument response function $\chi(\tau) = \exp(-\tau^2/\Delta\tau^2)$, with temporal response $\Delta\tau$. Furthermore, we account for the finite coherence time of the laser by including an exponential decay, such that the measured first-order correlation function becomes:
\begin{equation}
g_m^{(1)}(\tau) = e^{-\tau^2/\Delta\tau^2} e^{-\mu\vert\tau\vert} g^{(1)}_\mathrm{opt}(\tau),
\label{eq:SupExpg1}
\end{equation}
where $\tau_L = 1/\mu$ is the coherence time of the laser, and we have neglected any normalisation factors.

If we assume that we are operating deep inside the weak-driving regime, such that, $g^{(1)}_\mathrm{opt}(\tau)\approx g_\mathrm{coh}$, we can Fourier transform $g^{(1)}_m$ to extract the measured zero phonon line (ZPL) spectrum, neglecting the phonon sideband contribution.
This results in a Voigt profile with a Gaussian full width at half maximum through, $\mathrm{FWHM} = 4\ln 2 ~/\Delta\tau$.
Note that we have ignored the cavity contribution here. This is because, in the regime of interest, the cavity lineshape does not vary appreciably over the frequency range of the ZPL and therefore simply leads to a constant renormalisation of the spectra. 
We can now fit this Voigt profile to the measured ZPL, allowing us to match the theoretical model to the experiment according to Eq. \ref{eq:SupExpg1}.

\subsection{Dephasing and spectral wandering for off-resonant driving}

\begin{figure}
    \centering
    \includegraphics{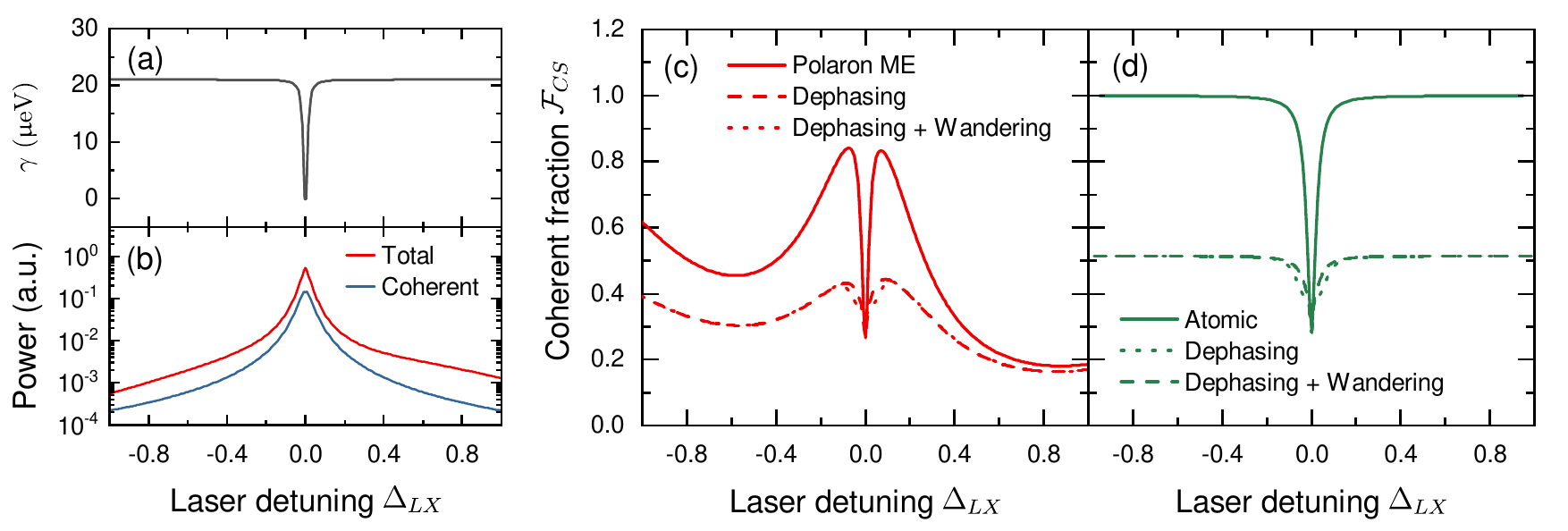}
    \caption{Dephasing and spectral wandering for off-resonant driving: (a) Plot of the extracted detuning-dependent dephasing rate $\gamma(\Delta_{LX})$. (b) Plot of the total and coherent powers calculated by the Polaron model. (c) Plot of the coherent fraction $\mathcal{F}_{CS}$ predicted by the Polaron model (solid line). The effects of including pure dephasing ($\gamma(\Delta_{LX})$) and both pure dephasing and spectral wandering are shown by the dashed and dotted lines respectively. (d) As for (c) but for the atomic model.}
    \label{fig:SupDetuning}
\end{figure}

When driving off-resonance as in Fig. 3 of the main text, we find that the polaron model tends to over-estimate the coherent fraction compared to the experiment. We note that excitation-induced dephasing due to phonons is already included within our model and thus propose that this discrepancy is due to additional pure dephasing that originates from charges generated by absorption of the laser in the doped bulk material \cite{doi:10.1063/1.321330}. As the behaviour for resonant driving is well-reproduced without this term (see Fig. 2 of main text), we suggest that the effect only becomes significant at larger laser detunings where the QD absorption cross-section reduces. To capture this within our model we introduce a detuning-dependent dephasing term:
\begin{equation}
    \gamma(\Delta_{LX}) = \gamma_\mathrm{max}(1-\xi^2/(\Delta_{LX}^2 + \xi^2)),
\end{equation}
with the linewidth $\xi$ fixed as the QD natural linewidth, representing the reducing QD absorption cross-section. Fitting the Polaron model with this additional term to the experimental data gives an amplitude $\gamma_\mathrm{max} = 21~\upmu\mathrm{eV}$. A plot of $\gamma(\Delta_{LX})$ is shown in Fig. \ref{fig:SupDetuning}(a) whilst Figs. \ref{fig:SupDetuning}(c,d) compare the Polaron (red lines) and atomic (green lines) models with (dashed lines) and without (solid lines) this pure dephasing term.

Compared to our experiments, the dip centered on resonance ($\Delta_{LX}^2 = 0$) produced by the Polaron calculation is somewhat too narrow. We suggest that this discrepancy arises from spectral wandering (on a timescale $\gg T_2$) in the sample that also likely originates from the charge environment surrounding the QD. Previous measurements on this sample observed a modest drop in two-photon interference visibility consistent with spectral wandering on a timescale of tens of nanoseconds \cite{Liu2018}. To account for these processes, we consider in Fig. \ref{fig:SupDetuning}(b) the total ($\mathcal{P}_{tot}$ - red line) and coherent ($\mathcal{P}_{coh}$ - blue line) powers emitted by the QD as calcualted by the Polaron model. Spectral diffusion causes the QD to sample these functions according to the distribution of the noise. Thus, after convolving these functions with an appropriate noise distribution, the coherent fraction $\mathcal{F}_{CS}$ may be evaluated according to:
\begin{equation}
    \mathcal{F}_{CS}(\Delta_{LX}) = \frac{\mathcal{P}_{coh}(\Delta_{LX})}{\mathcal{P}_{tot}(\Delta_{LX})}.
\end{equation}
As the power distributions are sharly peaked at $\Delta_{LX}=0$, the convolution has little effect when driving resonantly - any wandering results in a sharp drop in the emitted intensity and thus makes a very small contribution to the spectrum. However, away from resonance the gradient of $\mathcal{P}$ with $\Delta_{LX}$ is much smaller and the full wandering distribution is sampled. We are thus able to extract the wandering distribution by looking at the width of the incoherent peak in a spectrum with large $\Delta_{LX}$ such as Fig. 3(a) of the main text. Assuming a Gaussian noise distribution, fitting these spectra with a Voigt function with Lorentzian part fixed to the QD natural linewidth allows us to extract a FWHM of $66~\upmu\mathrm{eV}$ for the wandering distribution. The dotted lines in Figs. \ref{fig:SupDetuning}(c,d) show that the result of this convolution is to broaden the central dip of the calculated $\mathcal{F}_{CS}$ but with a negligible effect on the resonant case, consistent with the results of both Figs. 2 and 3 of the main text. We emphasize that the comparison between Figs. \ref{fig:SupDetuning}(c) and (d) illustrates that the additional dephasing and spectral wandering terms do not change the qualitative features of either the Polaron or atomic models.

\makeatletter\@input{xx.tex}\makeatother